\documentclass[twocolumn,amsmath,amssymb,aps,prb]{revtex4-2}
\usepackage{graphicx}
\usepackage{dcolumn}
\usepackage{bm}
\usepackage{vcell}
\usepackage{hyperref}
\usepackage{lipsum}
\usepackage{multirow}
\usepackage{color}
\begin{document}

\preprint{APS/123-QED}
\title{
 Tuning electronic properties in transition metal dichalcogenides MX$_2$ (M= Mo/W, X= S/Se) heterobilayers with strain and twist angle
}
\author{Ravina Beniwal$^{a}$, M. Suman Kalyan$^{b}$,  Nicolas Leconte$^{c}$, Jeil Jung$^{c, d,*}$, Bala Murali Krishna Mariserla$^{a,*}$, S. Appalakondaiah$^{a,e,*}$}
\affiliation{\small \it $^{a}$Department of Physics, Indian Institute of Technology Jodhpur, Rajasthan, India-342037}
\email{bmkrishna@iitj.ac.in}
\email{kondaiah@pondiuni.ac.in}
\affiliation{\small \it $^{b}$Department of Physics, National Institute of Technology Agartala, Agartala - 799046, Tripura, India.}
\affiliation{\small \it $^{c}$Department of Physics, University of Seoul, Seoul 02504, Korea}
\affiliation{\small \it $^{d}$Department of Smart Cities, University of Seoul, Seoul 02504, Korea}
\email{jeiljung@uos.ac.kr}
\affiliation{\small \it $^{e}$Department of Physics, Pondicherry University, Kalapet, Puducherry 605 014, India}
\date{\today}
\begin{abstract}
We explore the direct to indirect band gap transitions in MX$_2$ (M= Mo/W, X= S/Se) transition metal dichalcogenides heterobilayers for different system compositions, strains, and twist angles based on first principles density functional theory calculations within the G$_0$W$_0$ approximation. The obtained band gaps that typically range between 1.4$-$2.0 eV are direct/indirect for different/same chalcogen atom systems and can often be induced through expansive/compressive biaxial strains of a few percent. A direct to indirect gap transition is verified for heterobilayers upon application of a finite 16$^{\circ}$ twist that weakens interlayer coupling. The large inter-layer exciton binding energies of the order of $\sim$~250~meV estimated by solving the Bethe-Salpeter equation suggest these systems are amenable to be studied through infrared and Raman spectroscopy.

\end{abstract}
\maketitle

\section{Introduction}
\label{introduction}
Among the wide range of 2-dimensional (2D) materials, transition metal dichalcogenides (TMD) are promising due to their valley-dependent electronic structure ~\cite{Schaibley2016, Ominato2020}, strong excitonic (electron-hole pair) effects~\cite{Hill2015, Alexey2014, Ugeda2014}, 
and enhanced photoluminescence (PL) that make these ultra-thin materials suitable for novel flexible optoelectronic devices~\cite{ Mueller2018, Pospischil2016}. 
The artificial stacking of 2D layers as homo/hetero-bilayers 
allows to tune the optoelectronic properties
resulting in proposals of 
broadband photodetectors and flexible atomically thin p-n junction transistors~\cite{Zeng2018, Cheng2018, Liu2016, Zhou2018}. 
When van der Waals (vdW) homo/heterobilayers are composed of semiconductors, they have staggered band alignment which facilitates interlayer excitons due to charge transfer across the interfaces~\cite{Kozawa2016, Torun2018, Jiang2021, calman2018, Karni2019}. 
The exciton charges are spatially separated in different layers of TMD heterobilayers and hence endow them with long lifetimes lasting from hundreds of nano to micro seconds~\cite{Miller2017, Jiang2018}. 
The spatial charge separation between electron-hole pairs induces a permanent dipole moment, allowing electrically controlled optical and transport properties \cite{Jauregui2019}. These are systems with binding energies greater than 100~meV, making them suitable for developing devices that operate at room-temperature~\cite{ Liu2019}. 

Generally, the binding energies are lower for interlayer excitons in contrast to intralayer excitons. Particularly, monolayer TMD such as MX$_2$ (M= Mo, W, and X = S, Se), have exceptionally high exciton binding energies (320-700 meV)~\cite{Hill2015, Alexey2014, Liu2015}, while these energies are reduced in bilayers [80-450 meV] due to dielectric screening~\cite{Liu2015, Jo2014, Gerber2019}. 
Previous studies on the same chalcogen-centric heterobilayers MoS$_2$/WS$_2$ and MoSe$_2$/WSe$_2$ having very small lattice mismatch (0.3\%) displayed their indirect band gap behavior in the near-IR region with interlayer exciton binding energies of 250-450 meV respectively  \cite{Torun2018, Gillen2018}. The experimental studies based on TMD monolayers throw some light on dark excitons, which can interact with bright excitons and allow them to carry energy and information \cite{Madeo2020, Robert2020, Molas2017}.  The investigation of valley-polarized dark excitons in MoSe$_2$/WSe$_2$ reveals microsecond lifetimes, making them promising candidates for valleytronic devices and quantum computers~\cite{Jiang2018}.

Currently, most of the theoretical studies on TMD heterobilayers are focused on electronic, optical, and vibrational properties of same chalcogen atom monolayers or bilayers 
~\cite{Khatibi2019, Gao2017, Liang2014, Sahin2013}. However, there are limited reports on the electronic properties of different chalcogen atom heterobilayers
~\cite{Amin2016, Kang2013, Li2017}. Since lattice mismatch is arising in the heterobilayers, the band offset and band gap nature plays a crucial role in defining the optoelectronic properties, which remain yet to be examined. In this work, we have explored the ground-state electronic properties of same and different chalcogen-based heterobilayers and studied the nature of the band gaps as function of strain and twist angle.   
In our calculations, we carried out a systematic investigation of the vibrational properties, electronic band gap, and intra/interlayer exciton behavior using first-principles calculations for MX$_2$ (M= Mo/W, X= S/Se) heterobilayers. 
We explore the effects of strains and twist angles which clearly show the influence of interlayer interactions and lattice mismatch on aforementioned materials. 

This paper is structured as follows: Section \ref{comp_methods} describes the computational methodologies used to study the heterobilayers. In section \ref{str_prop}, we present structural and vibrational properties of TMD heterobilayers, followed by electronic properties in section \ref{ele_prop} and the absorption spectra in section \ref{opt_prop}. Band gap tuning of these structures through mechanical strain and twist angle is shown in Section \ref{Bandgap tuning}. We conclude the paper by summarizing our results in section \ref{summary}.

\section{Computational details}
\label{comp_methods}
The ground state and lattice dynamic calculations have been performed using plane-wave pseudopotential method implemented in Quantum Espresso \cite{Giannozzi2020}. We have used LDA-CAPZ \cite{Perdew1981} approximation as an exchange-correlation function and the vdW interactions are captured with
the opt-vdW86B dispersion correction~\cite{Klimes2011}. 
We considered 120~Ry plane wave kinetic energy cutoff with $16 \times 16 \times 1$ k-point grid in the Brillouin zone and added a vacuum height of 12~{\AA} to relax all heterobilayers. The Kohn–Sham self-consistent total energy was converged within 1E-8~eV. 
For both lattice and atomic relaxations, the positions of the atoms were allowed to relax until the forces were less than 1E-4~eV/\AA. To compute the Raman frequencies, we applied DFPT method with norm-conserving pseudopotentials~\cite{Corso1993}.
It is well-known that the Kohn-Sham calculations in DFT usually result in a bandstructure with underestimated band gaps. 
We have thus used the G$_0$W$_0$ approximation given by Vienna Ab-initio Simulation Package (VASP) to get a quasi-particle electronic bandstructure~\cite{Hedin1965}.
To achieve convergence, we utilized 432 bands for the summation over bands in the polarizability and self-energy calculations. Polarizability matrices are calculated up to a cutoff of 400~eV. We have studied excitonic properties by solving Bethe-Salpeter equation (BSE) ~\cite{Leng2016}. We considered the highest occupied valence and lowest unoccupied conduction bands as 18 and 22, respectively. The tunability of bandstructure under biaxial strains were calculated by applying constrained strains with specific ratios $x \% =(a-a_0)/a_0 \times 100 \% $, where $a$ and $a_0$ are the lattice parameters of the strained and unstrained structures, respectively. The unstrained optimized unit cell was enlarged or compressed symmetrically through different strain values. We have also verified the electronic properties of designed structures under twist to examine the nature of band gap. 

\section{Results and Discussion}\label{results}
\subsection{Structural and vibrational properties}\label{str_prop}
\begin{figure}[!ht]
    \centering
    \includegraphics[width=\linewidth]{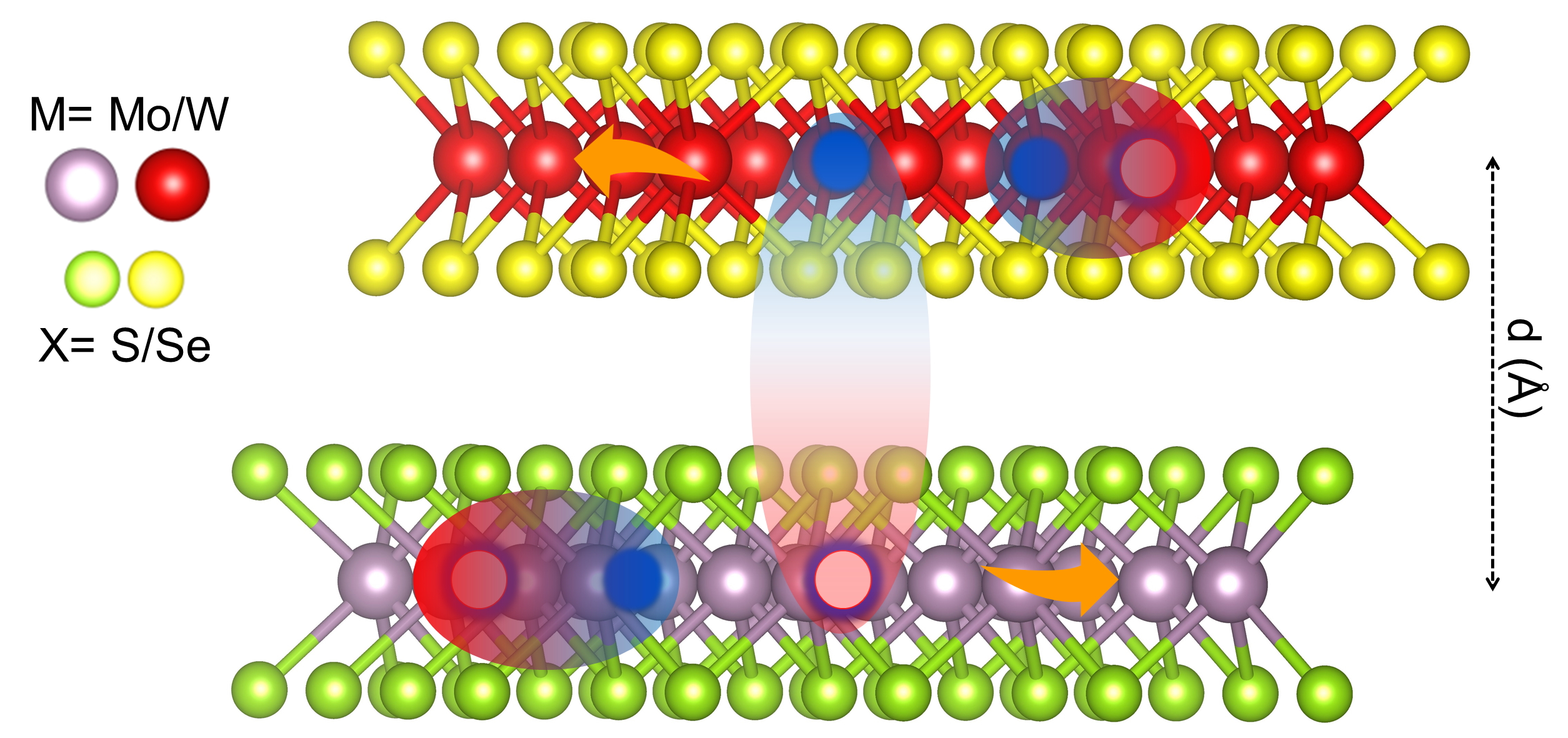}
    \caption{Schematic representation of a S/Se-based TMD heterobilayer with intralayer and interlayer exciton representations. Excitons in heterobilayers composed of two monolayers are intralayer, comprising electrons and holes in the same monolayer, and interlayer, comprising electrons and holes in subsequent monolayers.}
     \label{Structure}
\end{figure}

Our calculations began with stable AA$'$ stacking (chalcogen atoms of one layer overlap with the metal atoms of another layer and vice versa) in the heterobilayers in order to relax their geometries \cite{He2014, Li2016}. The ground state of studied heterobilayers is obtained by minimizing forces on both lattice and atomic positions. The resultant vdW-corrected lattice parameters and interlayer distances are listed in Table~\ref{T1}, which are averaged values of experimental bulk parameters. We have observed that these TMD-based heterobilayers (as illustrated in Fig.~\ref{Structure}) have inversion symmetry breaking and belong to {\it P3m1} space group with C$_{3v}$ symmetry, which is different from homobilayers having inversion D$_{3d}$ symmetry with {\it P$\bar{6}$m2} space group. This kind of symmetry breaking leads to novel and exceptional features for example in moiré structures~\cite{Jung2015}.

\begin{table*}
\caption{ Optimized lattice parameters (in \AA) (a), Interlayer distance (in \AA) (d), vibrational frequencies ($\omega$, in cm$^{-1}$) of MoS$_2$/MoSe$_2$, MoS$_2$/WSe$_2$, WS$_2$/MoSe$_2$, WS$_2$/WSe$_2$, MoS$_2$/WS$_2$, and MoSe$_2$/WSe$_2$. All the vibrational frequencies are both IR and Raman active. The experimental values are taken from \cite{Lin2015, Ye2019, MSKim2016, Zhang2016, Choudhary2016, Fu2018}} \label{T1}
\begin{ruledtabular}
\begin{tabular}{cccccccccccccccccccccccccccccc}
Structure     &a      &d        & $\omega$ (present)  &{$\omega$ (exp.)} & Mode     \\ \hline \\
 &  & & 17.7, 26.7 &  &  $E$(Shear), $A_1$(LBM)\\
 & & & 166.3, 237.8 & $\sim$  250 & $E$, $A_1$ (Se atom)\\
& &   & 272.8   &  &  $E$(S atom)  \\
 MoS$_2$/MoSe$_2$ & 3.221& 6.333   &  284.9, 350.2 &  &  $E$, $A_1$ (MoSe$_2$ unit)  \\
& &   &  365.3  & $\sim$  385 & $E$(MoS$_2$ unit)  \\
& &   &  393.4  & $\sim$  410 & $A_1$(S atom)  \\
 & & &  449.7 &   &  $A_1$ (MoS$_2$ unit)\\ \\ 
 
&  &  & 18.4, 25.4 &  & $E$(Shear), $A_1$(LBM)\\
& & & 169.3, 240.5 & & $E$, $A_1$ (Se atom)\\
& & & 241.2 & $\sim$  250 & $E$(WSe$_2$ unit)   \\
 MoS$_2$/WSe$_2$ &3.228 & 6.348 & 272.1&  &  $E$(S atom)      \\
& & & 301.5& & $A_1$(WSe$_2$ unit) &\\
& & & 364.2 & $\sim$  383  &$E$(MoS$_2$ unit) \\
& & & 393.2 & $\sim$  404  & $A_1$(S atom) \\
& & & 448.8 &  & $A_1$(MoS$_2$ unit) \\ \\
 
 &   &  & 18.5, 22.8 &  & $E$(Shear), $A_1$(LBM) \\
 & & & 166.2, 237.8 &  & $E$, $A_1$ (Se atom) \\
 WS$_2$/MoSe$_2$& 3.225 & 6.364 & 279 &  & $E$(S atom) \\
 & & &  284.8, &  & $E$(MoSe$_2$ unit) \\
 & & & 333.7 & & $E$(WS$_2$ unit) \\
 & & & 350.1& & $A_1$(MoSe$_2$ unit)  \\ 
  & & & 397.2 & & $A_1$(S atom)  \\ 
 & & & 412.8 & & $A_1$(WS$_2$ unit)  \\ \\
 
 &  &  &16.5, 20.3 &  &  $E$(Shear), $A_1$(LBM)\\
 & & & 169.3, 240.6 &  & $E$, $A_1$ (Se atom) \\
 WS$_2$/WSe$_2$  &3.232 & 6.379 & 241.1 & $\sim$ 250  &  $E$(WSe$_2$ unit) \\
 & & & 279.2 & $\sim$ 260 & $E$(S atom)   \\
  & & & 301.3 &  & $A_1$(WSe$_2$ unit)  \\
 & & & 332.7 & $\sim$ 355 & $E$(WS$_2$ unit) \\
 & & & 397.0 & $\sim$ 420 & $A_1$(S atom) \\ 
  & & & 411.8 &  & $A_1$(WS$_2$ unit) \\ \\
 
  &  &  & 19.3, 26.1 & $\sim$  31 &  $E$(Shear), $A_1$(LBM) \\
 MoS$_2$/WS$_2$  &3.162 & 6.177 & 278.5, 285.4 &  &  $E$ (S atom of both MoS$_2$, WS$_2$ units) \\
 & & & 342.5, 375.5 & $\sim$ 350, $\sim$ 375 & $E$ (WS$_2$ unit, MoS$_2$ unit)    \\  
 & & & 396.6, 400.5 & $\sim$ 410 & $A_1$ ( S atom of both MoS$_2$, WS$_2$ units)  \\
 & & & 422.8, 459.3 &  & $A_1$ (WS$_2$ unit, MoS$_2$ unit)  \\ \\

  &  &  & 16.6, 22.1 &  &  $E$(Shear), $A_1$(LBM) \\
 MoSe$_2$/WSe$_2$  & 3.296 & 6.534 & 164.1, 167.7 &  &  $E$ (Se atom of both MoSe$_2$, WSe$_2$ units) \\
 & & & 236.4 & $\sim$ 237 & $A_1$(Se atom of MoSe$_2$ unit)   \\  
 & & & 237.6 & $\sim$ 250 & $E$(WSe$_2$ unit) \\
 & & & 239.1 & $\sim$ 250 & $A_1$(Se atom of WSe$_2$ unit) \\ 
 & & & 279.6 & $\sim$ 287  & $E$(MoSe$_2$ unit) \\
 & & & 296.2, 344.7 & & $A_1$ (WSe$_2$ unit, MoSe$_2$ unit) \\
 \end{tabular}
\end{ruledtabular}
\end{table*}

\begin{figure*}[]
\centering
\includegraphics[width=\linewidth]{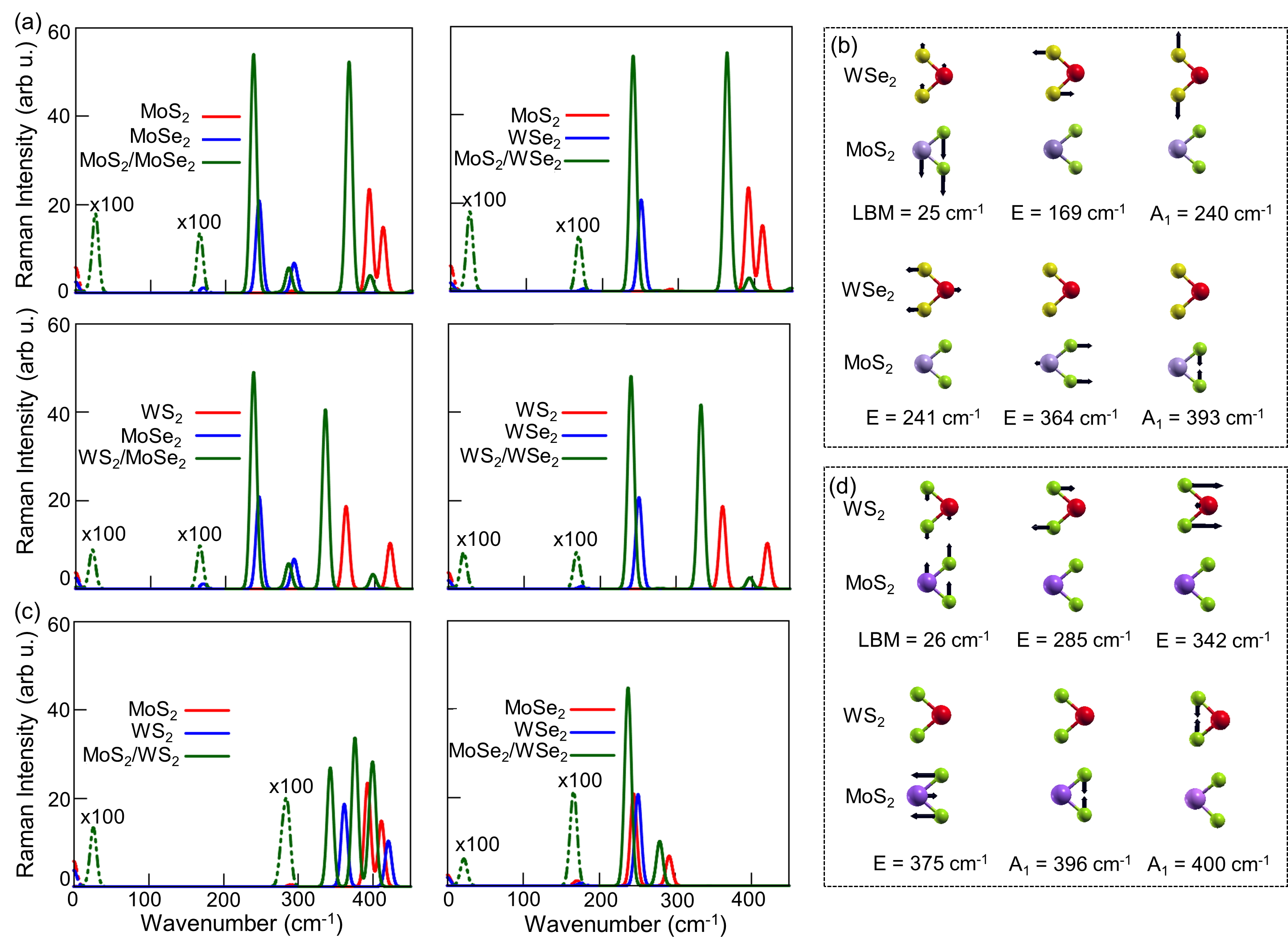}
\caption{(a) Computed Raman spectra for  MoS$_2$/MoSe$_2$, MoS$_2$/WSe$_2$, WS$_2$/MoSe$_2$ and WS$_2$/WSe$_2$ heterobilayers, (b) Illustration of the vibrational modes of MoS$_2$/WSe$_2$ heterobilayer. (c) Computed Raman spectra for MoS$_2$/WS$_2$ and MoSe$_2$/WSe$_2$ heterobilayers, (d) Illustration of the vibrational modes of MoS$_2$/WS$_2$ heterobilayer. From this, we can see the displacement of both layers in LBM modes.}
\label{Raman}
\end{figure*}

For a better understanding of interlayer interactions,
we have calculated their vibrational properties at the $\Gamma-$point. The interlayer interactions influence the atomic environment and create attractive or repulsive dipole forces that in turn affect the vibrational modes. A TMD-based heterobilayer has two units of X–M–X layers with C$_{3v}$ symmetry, implying that there are 18 phonon modes, out of which 3 are acoustic ($2E\oplus 1A_1$) and 15 optical ($10E\oplus 5A_1$) involving both in-plane ($E$) and out-of-plane ($A_1$) vibrations. All these modes are both Raman and IR active. The Raman spectra of heterobilayers and their monolayers are presented in Fig.~\ref{Raman}(a, c), and distinct fingerprints of different vibrational modes of same and different chalcogen-based heterobilayers are depicted in Fig.~\ref{Raman}(b, d).  
The modes in between 150-250 cm$^{-1}$ $\&$ 390-400 cm$^{-1}$ are from chalcogen atoms, and other modes are due to the metal with chalcogen atoms. We have summarized all the frequency modes and displacements for the studied heterobilayers in Table~\ref{T1}. The Raman intensities are magnified approximately twice for the heterobilayers over individual monolayers, which clearly indicates inter-layer interactions. The modes above $\sim$150 cm$^{-1}$ correspond to intralayer vibrations of each monolayer present in the heterobilayers. Importantly, we have observed two low-intensity modes below 50~cm$^{-1}$ region, these are due to lateral and vertical dipole-dipole interactions between the two layers in the heterobilayers~\cite{Tan2012, Zhao2013, Zhang2013, Lui2015}. These low-frequency shear and layer-breathing modes (LBM) have low intensity due to $0K$ calculations and can be intensified by raising the temperature. 

\subsection{Electronic properties}\label{ele_prop}
\begin{figure*}[!ht]
    \centering
    \includegraphics [scale=0.7]{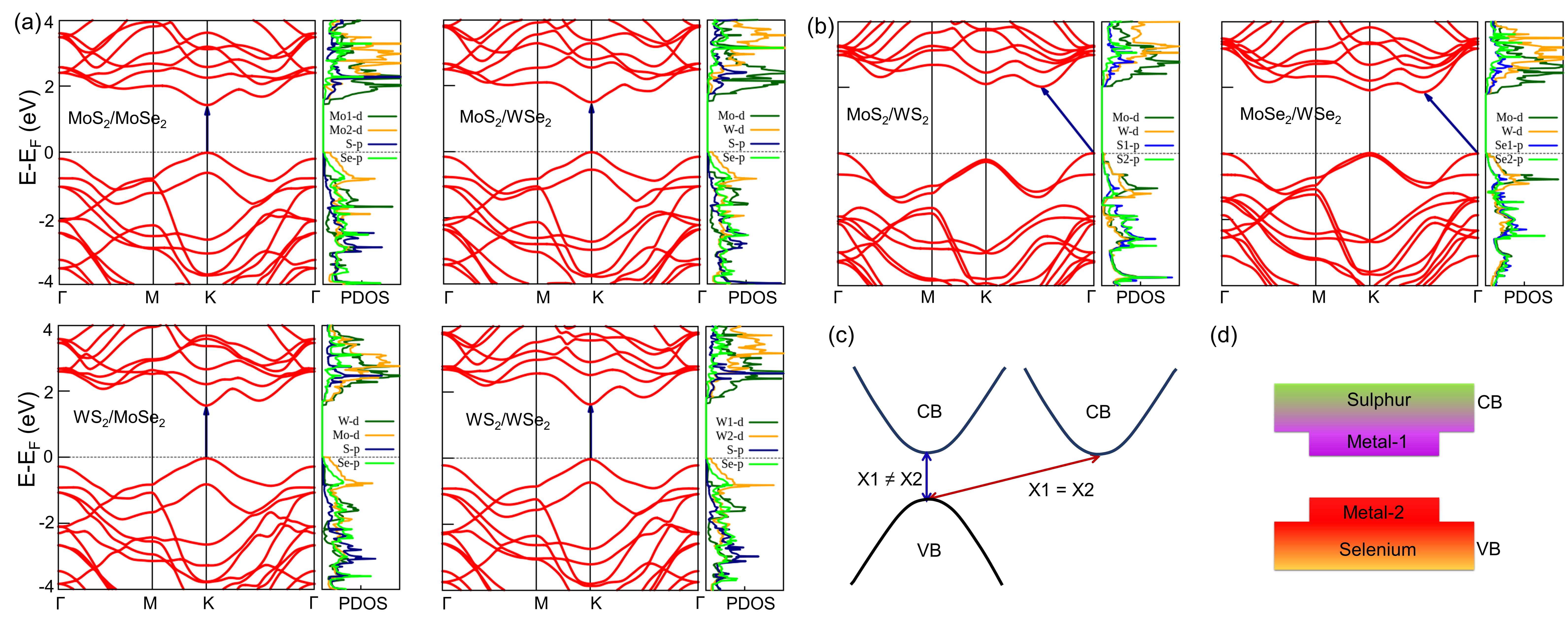}
    \caption{(a) Electronic bandstructure  and projected DOS  of different S/Se based heterobilayers using LDA + G$_0$W$_0$  approximation. All the compounds show direct band gaps.  (b). Electronic bandstructure and projected density of states of same S/Se based heterobilayers using LDA + G$_0$W$_0$  approximation. (c) Illustrative representation of direct and indirect band gap behavior based on the chalcogen atoms in MX$_2$ (M= Mo/W, X= S/Se) based heterobilayers. (d) Illustrative representation of the staggered band alignment of these heterobilayers.}
    \label{bandstructure}
\end{figure*}
The exact band gap calculation is of primary importance for optoelectronic device performance. It is known that the standard DFT (LDA/GGA) approaches underestimate the electronic band gap owing to the lack of derivative discontinuity in their exchange-correlation functional associated with the many-body electron-electron interactions. We have accounted for these effects using the G$_0$W$_0$ approximation, which provides accurate electronic properties for a wide range of materials~\cite{Appalakondaiah2013, Appalakondaiah2015}. 
The computed electronic bandstructure along with the orbital projected density of states (DOS) for different chalcogen-based heterobilayers are shown in Fig.~\ref{bandstructure}(a), and they show a direct band gap behavior along K-K high symmetry direction dissimilar to the indirect nature of the same chalcogen atom based heterobilayer (MoS$_2$/WS$_2$ \& MoSe$_2$/WSe$_2$) as shown in Fig.~\ref{bandstructure}(b). The band profiles of same chalcogen-based materials are consistent with previous reports, with a slight difference in the band gap values due to the absence of the spin-orbit coupling, which is responsible for lifting the degeneracy in the conduction and valence bands~\cite{Torun2018, Debbichi}. Direct/indirect band gap in a material depends upon the orbitals involved at the band edges. For different chalcogen-based materials, the transition metal d$_x$$_y$ orbitals have major contributions at both band edges and result in a direct band gap. For same chalcogen-based materials, chalcogens p$_z$ orbitals are more dominant at the valence band edge, and the transition metal d$_x$$_y$ orbitals populate the conduction band edge, which in turn changes the gap from direct to indirect. These results suggest that the material will have a direct band gap if X$_1$ $\neq$ X$_2$ otherwise indirect band gap, as illustrated in Fig.~\ref{bandstructure}(c).

Apart from the nature of the band gap, the band alignment (straddling/staggered/broken) of heterobilayers depends on the band offset of the individual monolayers.
We observed type-II band (staggered) alignment with band gap values in the range of 1.4 to 2.0 eV for the heterobilayers considered. The band gap value of MoS$_2$/WSe$_2$ is consistent with the reported experimental results ~\cite{Chiu2014} while the other heterobilayers require experimental verification. 
The obtained band gap values of heterobilayers are generally lower than those of TMD monolayers~\cite{Kim2021} and lie in the near IR region, which makes them useful for infrared applications. In the projected DOS, the valence band maxima is mainly coming from one layer (Se-based monolayer) and the conduction band minima from the other layer (S-based monolayer), as represented in Fig.~\ref{bandstructure}(d). At the band edges, the metal $d$-orbitals and chalcogen $p$-orbitals of each monolayer are highly dominating and the valence band maximum (conduction band minimum) has hybridization between metal-{$d$} orbital and Se (S)-{$p$} orbital resulting from the covalent bonds within each layer. 
\subsection{Optical properties}\label{opt_prop}
\begin{figure*}[!ht]
    \centering
    \includegraphics[width=\textwidth]{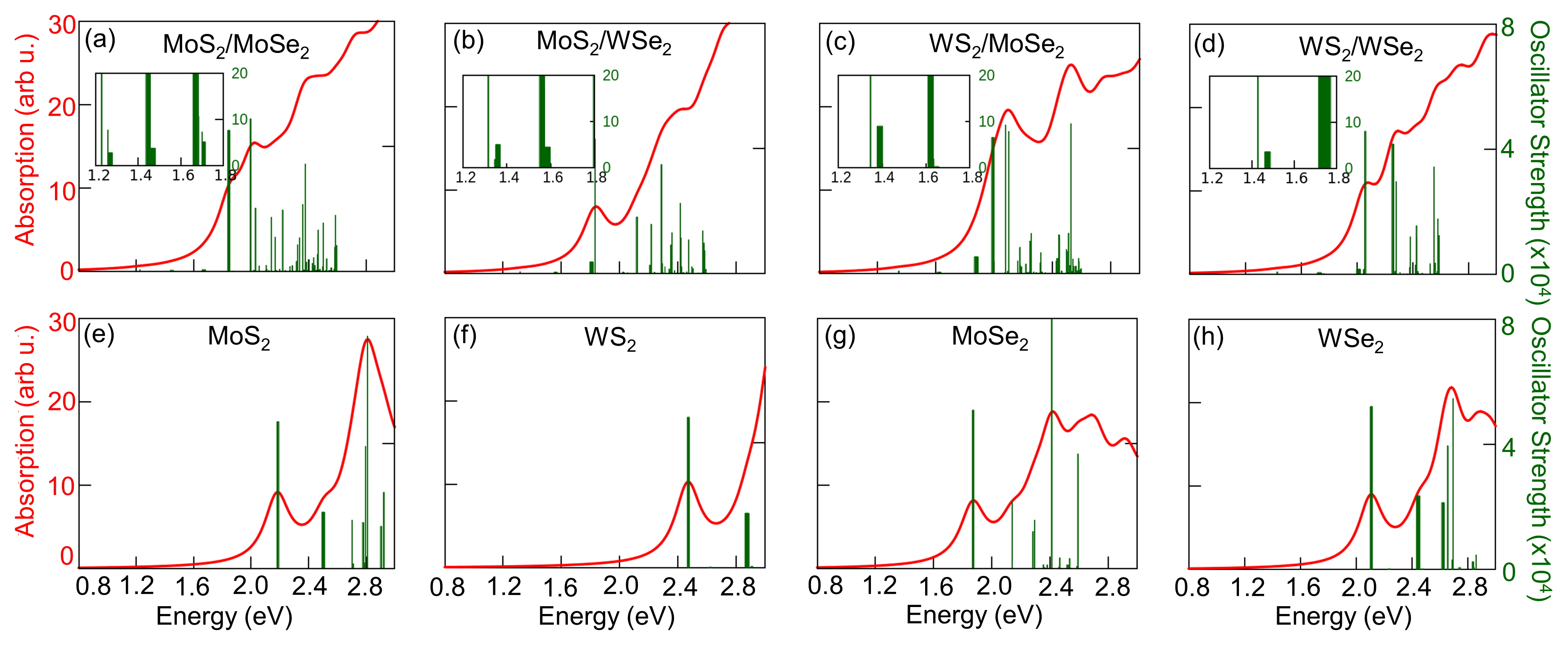}
    \caption{ Optical absorption spectra for (a) MoS$_2$/MoSe$_2$, (b) MoS$_2$/WSe$_2$, (c) WS$_2$/MoSe$_2$ and (d) WS$_2$/WSe$_2$ heterobilayer and independent monolayers (e) MoS$_2$, (f) MoSe$_2$, (g) WS$_2$ and (h) WSe$_2$ along with the oscillator strength. The insets illustrate the oscillator strength around the lowest interlayer excitonic peaks.} 
    \label{absorption}
\end{figure*}

Excitons of the same chalcogen TMD heterobilayers are well studied but the different chalcogen materials remain unexplored. Our current electronic studies show that different chalcogen heterobilayers have direct band gap and  the study of excitons in these materials help to understand the underlying features that control photoluminescence yield and the radiative recombination rate of electrons and holes. These layer-separated electron–hole interactions are examined through the BSE approximation~\cite{Leng2016} to estimate the optical absorption of hetero-bilayers (shown in Fig.~\ref{absorption}) along with the oscillator strengths. The optical band gaps are found in the range of 1.225 to 1.428~eV for heterobilayers, (as shown in Table \ref{T2}), which are interlayer exciton transitions having binding energies in the range of 240-250~meV with weak oscillator strengths due to weak dipole-dipole interactions between the layers. These interlayer excitons are quite stable from thermal dissociation and useful for developing stable optoelectronic devices~\cite{Lukman2020}. It is well known that monolayers show A and B intralayer excitons in the visible spectral region~\cite{Palummo2015, Ceballos2016, Zhu2015} with large binding energies, whereas the formation of excitons in different chalcogen-based heterobilayers lies in infrared spectral domain. 

The excitons can be tuned by selecting different metal atoms and by controlling the interlayer interactions. For instance, MoS$_2$/MoSe$_2$ has the lowest electronic band gap, the A and B exciton peaks are located at 1.850 (A$_1$), 1.997 (A$_2$), and 2.029 (B$_1$), 2.057 (B$_2$) eV, respectively, as shown in Table~\ref{T2}. When one of the metal atoms is replaced with W (MoS$_2$/WSe$_2$), the electronic band gap increases, and the peaks corresponding to A and B excitons shift to 1.820 (A$_1$), 1.831 (A$_2$) and 2.122 (B$_1$), 2.212 (B$_2$)~eV, respectively. These two sets of values correspond to the individual monolayers. By replacing the W atoms with Mo, we can maximize the band gaps in the studied heterobilayers and cause a slight shift in the A, and B exciton peaks. The localization of e$^-$ and hole pairs within the layer has high oscillator strengths, and their delocalization reduces oscillator strengths which are observed for interlayer excitons. The energetic positions of exciton peaks are slightly shifted compared to individual monolayers because of small dielectric screening along the out of plane direction. The current result closely matches with the reported experimental photoluminescence measurements of the heterobilayers \cite{Arora2021, Ye2019, Karni2019, Kozawa2016}.

 \begin{table*}
\renewcommand\arraystretch{1.3}
\begin{tabular}{|>{\centering\arraybackslash}p{2.2cm}|>{\centering\arraybackslash}p{1.8cm}|>{\centering\arraybackslash}p{1.8cm}|>{\centering\arraybackslash}p{1.8cm}|>
{\centering\arraybackslash}p{1.8cm}|>
{\centering\arraybackslash}p{1.8cm}|>
{\centering\arraybackslash}p{1.8cm}|}
\hline
{Material} & $E_g$ \hspace{1cm} (LDA) & {$E_g$ \hspace{1cm} (G$_0$W$_0$)}& {$E_g$ (Optical)} &  {A exciton}  & B exciton & {Binding energy } \\ \hline
$\text{MoS}_2$/$\text{MoSe}_2$ & 0.861 & 1.465 & 1.225 &  A$_1$-1.850, A$_2$-1.997 & B$_1$-2.029, B$_2$-2.057 & 0.240 \\ \hline
$\text{MoS}_2$/$\text{WSe}_2$ & 0.654 & 1.560 & 1.315  &  A$_1$-1.820, A$_2$-1.831  & B$_1$-2.122, B$_2$-2.212 & 0.245   \\ \hline
$\text{WS}_2$/$\text{MoSe}_2$ & 1.117 & 1.591 & 1.345 &  A$_1$-1.889, A$_2$-1.991 & B$_1$-2.078, B$_2$-2.100 & 0.246  \\ \hline
$\text{WS}_2$/$\text{WSe}_2$ & 0.946 & 1.679 & 1.428 &  A$_1$-2.049, A$_2$-2.055 & B$_1$-2.262, B$_2$-2.280 & 0.251 \\ \hline

\end{tabular}
\caption{ Obtained band gaps using LDA,  $G_0W_0$ approximations, Optical band gap, intralayer A and B exciton positions and exciton binding energies of the studied heterobilayers. All energies are in eV.}\label{T2}
\end{table*}
\subsection{Band gap nature under strain and twist } \label{Bandgap tuning}
\begin{figure*}[!ht]
    \centering
    \includegraphics 
    [scale=0.95]{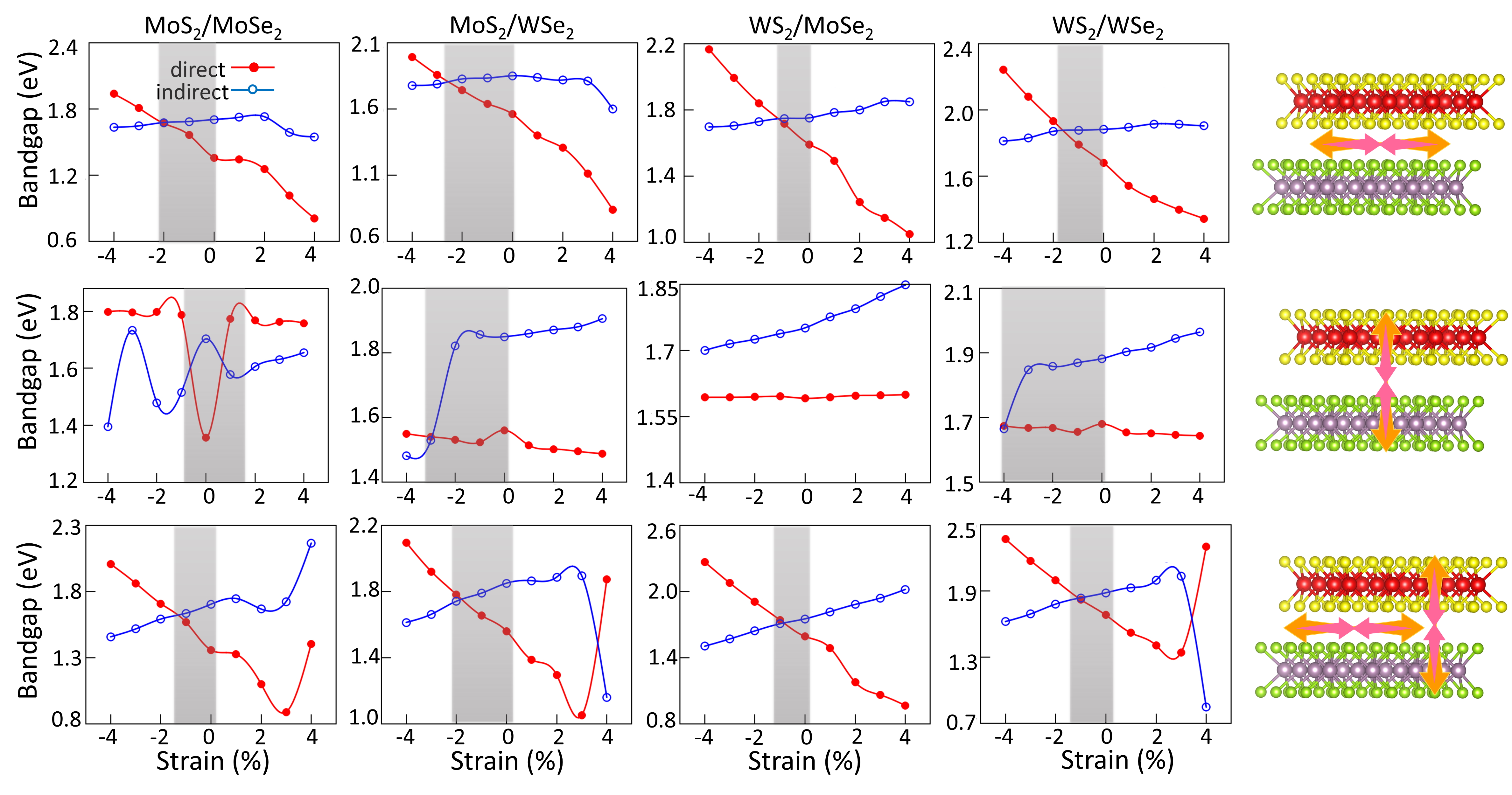}
    \caption {Effect of different constrained strains to examine the nature of the band gap behavior for TMD-based heterobilayers using LDA + G$_0$W$_0$  approximation, {\it{Top panel}}: In-plane strain with a fixed interlayer distance, {\it{Middle panel}}: Out of plane strain at a fixed in-plane lattice constant,  {\it{Bottom panel}}: Biaxial strain, where both in- and out-of-plane parameters are varied}
    \label{Strain}
\end{figure*}
 \begin{figure*}[!ht]
    \centering
    \includegraphics [scale=0.95]{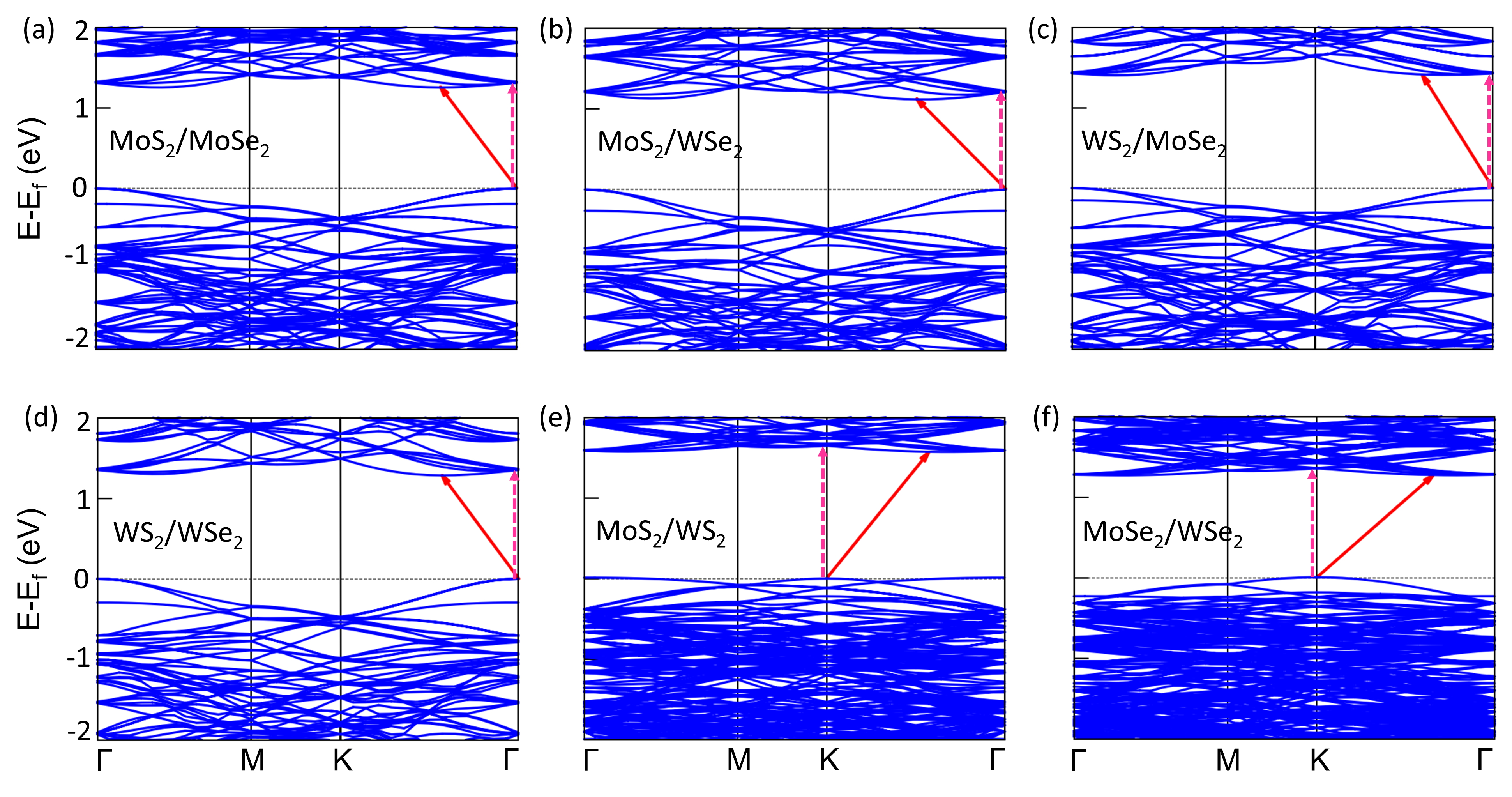}
    \caption{Calculated folded  bandstructure for 16$^{\circ}$ twisted heterobilayers using LDA approximation for (a) MoS$_2$/MoSe$_2$, (b) MoS$_2$/WSe$_2$, (c) WS$_2$/MoSe$_2$, (d) WS$_2$/WSe$_2$, (e) MoS$_2$/WS$_2$, and (f) MoSe$_2$/WSe$_2$ (red line indicates indirect band gap and pink line represents direct band gap)}
    \label{Twisted_bandstructure}
\end{figure*}
In the previous sections, we evaluated the electronic and optical properties of structurally relaxed heterobilayers, where the lattice constant is average of both individual monolayers. 
The change in lattice parameter affects the fundamental properties as observed in earlier report on MoSe$_2$/MoS$_2$, shows the band gap behavior changing from direct to indirect~\cite{Arora2021}. This distinct behavior indicates an external strain acting on these materials plays a vital role in describing their electronic properties. In this aspect, we explore the electronic band gap behavior by applying mechanically constricted strain to the relaxed geometries. Here, the strain is applied in-plane, out-of-plane and both together. As shown in Fig.~\ref{Strain}, a minimal strain changes the band gap and its nature along the momentum direction. By applying in-plane compression, we observed the smooth change in the band gap from direct to indirect, and for the out-of-plane compressions, the band gap changes occurred asymptotically. The out-of-plane strain (vdW interactions) influence the band gap nature between 2-4{\%} applied strain in MoS$_2$/WSe$_2$, WS$_2$/WSe$_2$ and below 1{\%} external strain for MoS$_2$/MoSe$_2$. But WS$_2$/MoSe$_2$ doesn't change the band gap nature within 4{\%} of strain. Whereas, the simultaneous in and out-of-plane compression changes the band gap nature within 2{\%} of the compression strain, effectively due to the dominance of in-plane interactions. These results opens the door to tune the band gap and its nature from direct to indirect and vice versa by selective strain in TMD hetrobilayers through experiments.

In addition, the lattice mismatch in the individual layers plays a crucial role in forming the commensurate structures for the TMD heterobilayers. For instance, the same chalcogen-based monolayers (M$_1$ $\neq$ M$_1$, X$_1$ = X$_2$) possess a lattice mismatch of 0.3{\%}, and different chalcogen-based layers (X$_1$ $\neq$  X$_2$)  have a lattice mismatch of 3.8{\%}. 
 Recent report on MoS$_2$/WSe$_2$ commensurate structure with a 16$^{\circ}$ twist angle exhibited both direct and indirect band gap behavior~\cite{Karni2019}, flagged the importance of twist in TMD heterobilayers. Hence, we examined the 16$^{\circ}$ twist angle heterobilayers in the present work, and {these twisted supercells are generated by simulating the structure with the twister code~\cite{Naik2018}. The new supercell lattice vectors (called X and Y) are related to the lattice vectors of each monolayer through the following matrices: 
\begin{gather}
 \begin{bmatrix} X  \\ Y 
 \end{bmatrix}
 =
 T
  \begin{bmatrix}
   x_1 (M_1X_1)  \\
   y_1 (M_1X_1)
   \end{bmatrix} 
    =
 T'
  \begin{bmatrix}
   x_2 (M_2X_2)  \\
   y_2 (M_2X_2)
   \end{bmatrix}
\end{gather}
where $x_1$($M_1$$X_1$), $y_1$($M_1$$X_1$), $x_2$($M_2$$X_2$) and $y_2$($M_2$$X_2$) are the lattice vectors of each monolayer respectively, T and  T$^{\prime}$ are transformation matrices given by
\begin{equation}
    T=\begin{pmatrix}
  3 & -4   \\
   4 & -1
\end{pmatrix}
{\rm and} ~ T' = \begin{pmatrix}
  2 & -4   \\
   4 & -2
\end{pmatrix}
\end{equation}

The supercell generated with these vectors contains a larger number of atoms, and dealing with such large structures using G$_0$W$_0$ methodology is computationally expensive. Thus, we calculated these band structures using the LDA approximation with 8×8×1 k-point mesh. The resultant folded electronic band structures are shown in Fig.~\ref{Twisted_bandstructure}. 
It is clearly evident that different (same) chalcogen-based heterobilayers shows the momentum-indirect band gap in between $\Gamma-K$ ($K-\Gamma$) and an energy difference of 10-100~meV with the momentum-direct band gap nature along the $\Gamma$(K)- high symmetry direction (band values are given in Table~\ref{T3}). Interestingly, we found that the bandwidth of valence and conduction bands are decreased from untwisted to twisted heterobilayers, suggesting the possibility of obtaining flat-bands with twist angle. These findings of TMD-heterobilayers have shown great possibilities to tune the electronic and optical properties with minimal twist and strain for optoelectronics and solar-cell device applications. 

\begin{table*}
\renewcommand\arraystretch{1.3}
\begin{tabular}{|>{\centering\arraybackslash}p{2.2cm}|>{\centering\arraybackslash}p{1.8cm}|>{\centering\arraybackslash}p{1.8cm}|>{\centering\arraybackslash}p{1.8cm}|>
{\centering\arraybackslash}p{1.8cm}|>
{\centering\arraybackslash}p{1.8cm}|>
{\centering\arraybackslash}p{1.8cm}|>
{\centering\arraybackslash}p{1.8cm}|}
\hline
{Material} & \multicolumn{2}{c|}{Without twisted geometry} & \multicolumn{2}{c|}{With twisted geometry} & Twisted geometry $\Delta$E$_(direct-indirect)$ (meV) & Without twisted geometry $\Delta$E$_(direct-indirect)$  (meV)\\  \cline{2-5}
& VB (eV) & CB (eV) & VB (eV) & CB (eV) &\\   \hline
$\text{MoS}_2$/$\text{MoSe}_2$ & 0.853 & 0.978 & 0.432 &  0.155 & 57 & 255 \\ \hline
$\text{MoS}_2$/$\text{WSe}_2$ & 0.866 & 0.999 & 0.619 &  0.155 & 92 & 336 \\ \hline
$\text{WS}_2$/$\text{MoSe}_2$ & 0.833 & 1.037 & 0.497 &  0.225 & 18 & 150 \\ \hline
$\text{WS}_2$/$\text{WSe}_2$ & 0.877 & 1.058 & 0.614 &  0.210 & 68 & 217 \\ \hline
$\text{MoS}_2$/$\text{WS}_2$ & 1.340 & 0.963 & 0.167 &  0.078 & 78 & 61 \\ \hline
$\text{MoSe}_2$/$\text{WSe}_2$ & 1.111 & 1.325 & 0.348 &  0.088 & 85 & 82 \\ \hline
\end{tabular}
\caption{Calculated linewidth of the valence and conduction bands without and with twist geometry. Direct to indirect band gap difference in twisted and relaxed structures}

\label{T3}
\end{table*}

\section{Conclusions}\label{summary}
In summary, we have explored the electronic and optical properties of MX$_2$ (M= Mo/W, X= S/Se) heterobilayers using first-principles density functional theory calculations. Designed heterobilayers of TMD-based materials 
exhibited inversion symmetry breaking with point group C$_{3v}$,  shows Raman fingerprints of individual monolayers are at higher frequencies ($>$ 150 cm$^{-1}$) and below 50 cm$^{-1}$ correspond to the shear or layer breathing modes due to the atomic environment \& interlayer interactions.
 We have used the G$_0$W$_0$ method to calculate accurate electronic properties and found that the electronic band gaps are in the range of 1.4 to 2.0~eV with staggered band alignment. We found that the band gap behavior changes from direct to indirect and vice versa by choice of the chalcogen (same/different) atoms in the heterobilayers.
 The absorption spectra are calculated using BSE methodology and obtained the exciton binding energies for different chalcogen-based heterobilayers are in the order of ~250 meV due to strong interlayer interactions. We also investigated the electronic bandstructures under the in-plane, out-of-plane strains $\&$ twist angle between heterobilayers and observed change in band gap nature and its values. Our results demonstrate that, under minimal strain, different chalcogen-based heterobilayers can change their band gap from direct to indirect owing to their interlayer interactions. These outcomes of TMD-heterobilayers have shown great possibilities to tailor the electronic and optical properties with minimal twist for tunable optoelectronic applications.\\

\begin{acknowledgments}\label{acknowledgments}
RB would like to thank UGC for the senior research fellowship and computational facilities at IITJ. This work is funded by Science and Engineering Research Board (SERB) grant number RJF/2021/000147 (SA), SERB:CRG/2022/008749 (BMKM), IITJ seed grant:I/SEED/BMK/20230017  (BMKM) and the Korean NRF through Grants with No. 2020R1A2C3009142 (N.L.). We are grateful to the computational facilities from KISTI Grant No. KSC-2022-CRE-0514 and the resources of Urban Big data and AI Institute (UBAI). J.J. acknowledges the Korean NRF grant with No. 2020R1A5A1016518 and the Korean Ministry of Land, Infrastructure and Transport (MOLIT) from the Innovative Talent Education Program for Smart Cities.

\end{acknowledgments}


%

\end{document}